\documentclass[prl,aps,twocolumn,showpacs]{revtex4}

\usepackage{epsfig}
\usepackage{amssymb}
\usepackage{graphicx}

\begin{document}

\title{Selective interactions in trapped ions: state reconstruction
and quantum logic}

\author{E. Solano}

\affiliation{Max-Planck-Institut f{\"u}r Quantenoptik,
Hans-Kopfermann-Strasse 1, D-85748 Garching, Germany\\
Secci\'{o}n
F\'{\i}sica, Departamento de Ciencias, Pontificia Universidad
Cat\'{o}lica del Per\'{u}, Apartado 1761, Lima, Peru}

\date{\today}

\begin{abstract}
We propose the implementation of selective interactions of
atom-motion subspaces in trapped ions. These interactions yield
resonant exchange of population inside a selected subspace,
leaving the others in a highly dispersive regime. {\it
Selectivity} allows us to generate motional Fock (and other
nonclassical) states with high purity out of a wide class of
initial states, and becomes an unconventional cooling mechanism
when the ground state is chosen. Individual population of number
states can be distinctively measured, as well as the motional
Wigner function. Furthermore, a protocol for implementing quantum
logic through a suitable control of selective subspaces is
presented.
\end{abstract}


\maketitle

The interaction of a two-level with an infinite-dimensional system
is one of the most simple and fundamental quantum models,
describing the interplay between discrete and continuous variable
systems. The case of a two-level atom interacting with a single
mode of the electromagnetic field, typically described by the
Jaynes-Cummings (JC) model~\cite{JC} in cavity quantum
electrodynamics (CQED)~\cite{WaltherReview,HarocheReview}, can be
reproduced with more complexity in the context of an ion trapped
in a harmonic potential~\cite{WinelandBlattReview}. In the latter,
two internal levels are coupled to a vibrational mode through an
adequately tuned laser system. CQED and trapped ion systems have
been studied along different research lines like the generation
and measurement of
nonclassical~\cite{HarocheCat,WinelandCat,ionsuperpositions,WaltherPhoton}
and/or entangled
states~\cite{WinelandEntanglement,HarocheEntanglement}, state
reconstruction~\cite{WinelandWigner,HarocheWigner} and the
implementation of quantum processing and computing
devices~\cite{HarocheLogic,WinelandLogic,BlattLogic}. Typically,
the interactions used for those purposes, in resonant or
dispersive regimes, involve dynamically all states of the
associated atom-field or atom-motion Hilbert space.

In this letter, we propose the implementation of selective
interactions in trapped ion systems. {\it Selectivity} is
associated with the possibility of producing resonant interactions
exclusively inside preselected atom-motion Hilbert subspaces,
while all others remain in a highly dispersive regime. We describe
how to tailor the interaction of a two-level system (like a
two-level atom) with an infinite dimensional system(like a
harmonic oscillator) into a resonant interaction of a reduced two
two-level systems. We show that, beyond its fundamental interest,
selectivity is a powerful tool for achieving diverse quantum
effects, from nonclassical state generation, cooling to the ground
state, state reconstruction, and quantum logic.

Selectivity, in the way it is presented here, can be related and
eventually applied to quantum effects like
blockade~\cite{blockade}, individual (selective)
addressing~\cite{individualopticallattice}, and
turnstile~\cite{turnstile} mechanisms, among others. Early
attempts to implement similar (selective) devices were developed
in recent years in
CQED~\cite{PellizariSelectivity,NhaSelectivity,CQEDselectivity}
and, quite specifically, in the collective behavior of two trapped
ions outside the Lamb-Dicke regime~\cite{TwoIonSelectivity}.

The interaction of a two-level system with a harmonic oscillator,
like atom-field (atom-motion) interactions in CQED (trapped ions),
can be described by the JC model, whose Hamiltonian under resonant
conditions and in the interaction picture reads
\begin{eqnarray}
\label{JCHamiltonian} H^{\rm I}_{\rm JC}=\hbar g (
{\sigma}^{\dagger} a + \sigma a^{\dagger}) .
\end{eqnarray}
Here, $a$ ($a^{\dagger}$) is the annihilation (creation) operator
associated with a harmonic oscillator and $\sigma$
($\sigma^{\dagger}$) is the lowering (raising) atomic operator $|
g \rangle \langle e |$ ($| e \rangle \langle g |$), where $| g
\rangle$ ($| e \rangle$) is the ground (excited) atomic state.
This Hamiltonian yields Rabi oscillations~\cite{Rabi} in all JC
subspaces
\begin{eqnarray}
\label{JCdoublets} \{ | g, n \rangle, | e, n - 1 \rangle \}
\end{eqnarray}
with $n=0,1, ...$, where $| n \rangle$ is a number state of the
harmonic oscillator.

It is possible to produce effectively anti-Jaynes-Cummings (AJC)
interactions in CQED~\cite{CQEDcats} and trapped
ions~\cite{IonJC}, whose Hamiltonian reads
\begin{eqnarray}
\label{AJCHamiltonian} H^{\rm I}_{\rm JC}=\hbar g (
{\sigma}^{\dagger} a^{\dagger} + {\sigma} a) ,
\end{eqnarray}
and produces Rabi oscillations in all AJC subspaces
\begin{eqnarray}
\label{AJCsubspace} \{ | g, n \rangle, | e, n + 1 \rangle \}
\end{eqnarray}
with $n=0,1, ...$.

We call {\it selective interaction} to a resonant interaction
producing exchange of population (a selective Rabi oscillation)
inside a chosen JC or AJC subspace,
\begin{eqnarray}
\label{JCsubspace} \{ | g, N_0 \rangle, | e, N_0 \pm 1 \rangle \}
,
\end{eqnarray}
with fixed $N_0$, while all other subspaces remain strictly
off-resonance.

Beyond the fundamental interest in this particular tailoring of
the Hilbert space, selectivity provides us with a flexible tool
that will prove to be useful in a wide range of applications.
Here, we will discuss its implementation in trapped ion systems,
although the different examples should be valid for any other
physical system where selectivity could be implemented.

For the sake of simplicity, we choose a simplified setup
consisting of a single trapped ion coupled to its unidimensional
center-of-mass (CM) motion through a bichromatic laser excitation.
These ideas can be generalized straightforwardly to many ions and
many (collective) motional modes. We consider a Raman laser
excitation of a three-level trapped ion as shown in Fig. 1. The
transitions between internal states $| g \rangle \leftrightarrow |
c \rangle$ and $| e \rangle \leftrightarrow | c \rangle$ are
excited off resonantly (large detuning $\Delta$) by a
standing-wave field with coupling strength $g_1$ and by a
travelling-wave field with coupling strength $g_2$, respectively.
The Raman scheme is realized in such a way that the effective
(resonant) two-photon process involves the excitation of the first
blue vibrational sideband through the $| g \rangle \leftrightarrow
| c \rangle$ transition. In the rotating wave approximation, the
Hamiltonian describing this system is
\begin{eqnarray}
H = && \hbar \nu a^{\dagger} a + \hbar \omega_e
| e \rangle \langle e |+ \hbar \omega_c | c \rangle \langle c |
\nonumber \\
[1ex] && + \hbar g_1 \bigg\lbrack e^{-i(k_1 z - \omega_1 t)} +
e^{i(k_1 z + \omega_1 t)} \bigg\rbrack | g \rangle \langle c |
\nonumber \\ && + \hbar g_2 e^{-i(k_2 z - \omega_2 t)} | e \rangle
\langle c |  + {\rm H. c.} ,
\end{eqnarray}
where the coupling strengths $g_1$ and $g_2$ are taken as positive
and $\omega_i = k_i c$ ($i = 1,2$) are the frequencies of the
exciting fields.
\begin{figure}[!t]
\includegraphics[width=55mm]{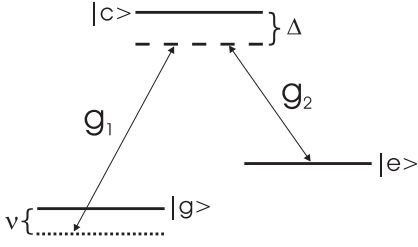}
\vspace*{0.0cm} \caption{\label{fig1} Energy diagram of the
three-level ion. $g_1$ ($g_2$) is the coupling strength of a
standing-wave (travelling-wave) field with the shown atomic
transitions, and $\nu$ is the trap frequency.}
\end{figure}
In the Lamb-Dicke (LD) regime~\cite{comment}, in the interaction
picture, and eliminating adiabatically level $| c \rangle$ under
the dispersive condition $\Delta \gg \Omega_{\rm eff} \equiv 2
\eta_2 g_1 g_2 / \Delta$, we can write the effective AJC-like
Hamiltonian
\begin{eqnarray}
\label{Heff} H_{\rm eff} = \!\!\!\!\! && -4 [ \hbar
\frac{g_{1}^{2}}{\Delta} - \hbar \eta_{1}^2
\frac{g_{1}^{2}}{\Delta} (2
a^{\dagger} a + 1) ] | g \rangle \langle g | \nonumber \\
&& - \hbar \frac{g_{2}^{2}}{\Delta} | e \rangle \langle e
| \nonumber \\[1ex] && + 2 i \hbar \eta_{2} \frac{g_{1} g _{2}}{\Delta}
( | e \rangle \langle g | \, a^{\dagger } - | g \rangle \langle e
| \, a) ,
\end{eqnarray}
where $\eta_{i} \equiv k_i \sqrt{\hbar / 2 m \nu}$ are the
 LD parameters corresponding to each laser field excitation.

Associated with the transition in the AJC subspace $\{ | g, N_0
\rangle , | e, N_0 + 1 \rangle \}$, the effective Hamiltonian in
Eq.~(\ref{Heff}), $H_{\rm eff}$, shows a detuning frequency
\begin{eqnarray}
\Delta^ {N_0}_{eg} = - 4 \eta_{1}^2 \frac{g_1^2} {\Delta} (2 N_0 +
1) + ( 4 \frac{g_1^2} {\Delta} - \frac{g_2^2} { \Delta } ) .
\end{eqnarray}
This detuning can be compensated for a fixed phonon number $N_0$
by DC Stark shift or by shifting the laser frequencies.
Selectivity appears when we tune to resonance a preselected
subspace transition $\{ | g, N_0 \rangle \leftrightarrow | e, N_0
+ 1 \rangle \}$, while all other AJC subspaces $\{ | g, n \rangle
, | e, n + 1 \rangle \}$, with $n \neq N_0$, remain dispersive.
Once the correction is done specifically for
$\{\left|g,N_{o}\right\rangle,\left| e,N_{o}+1\right\rangle\}$,
the remaining detunings associated with other subspaces ($n\neq
N_0$) are
\begin{eqnarray}
{\Delta^{n}_{eg}}^* \equiv \Delta^ {n}_{eg} - \Delta^{N_0}_{eg}= -
8 \eta_{1}^2 \frac{g_{1}^2} {\Delta} (n - N_0) .
\end{eqnarray}
If after this reshifting process, the dispersive condition
\begin{eqnarray}
| {\Delta^{n}_{eg}}^* | \gg | \Omega_{\rm eff} | \equiv 2 \eta_{2}
\frac{g_1 g_2}{\Delta}
\end{eqnarray}
holds $\forall n\neq N_0$, we arrive to the selectivity condition
\begin{eqnarray}
{\cal S} \equiv 4  \frac{\eta_{1}^2}{\eta_{2}} \frac{g_1}{g_2} \gg
1
\end{eqnarray}
for the selectivity parameter ${\cal S}$.

Considering the experimental parameters of the ion experiments at
NIST (Boulder)~\cite{WinelandEntanglement}, as an example without
optimization, and imposing strictly the selectivity condition,
${\cal S} \gg 1$, it is possible to achieve
\begin{eqnarray}
\Omega_{\rm eff} \equiv 2 \eta_2 \frac{g_1 g_2}{\Delta} \lesssim
10^5 {\rm Hz} .
\end{eqnarray}
This effective coupling strength produces population inversion of
any selected subspace $\{\left|g,N_0\right\rangle \leftrightarrow
\left| e,N_0+1\right\rangle\}$ in $\tau_{\rm inv} < 0.1$ ms. Then,
the required times for implementing selectivity in trapped ions
are shorter compared to the motional decoherence time, typically
$\tau_{\rm dec} \gtrsim 10$ ms. For sure, it will be also
interesting to design a selective scheme in other ion setups, like
the one in the Innsbruck group~\cite{WinelandBlattReview}, where,
at variance with the Raman scheme in NIST, a quadrupolar two-level
transition is directly and strongly excited. This interest is well
founded since recently F. Schmidt-Kaler {\it et
al.}~\cite{Innsbruckselectivity} have realized a 2-qubit gate
through an off-resonant coupling of a laser field to a motional
sideband, producing a phase shift conditioned to the motional
state and showing that a selective mechanism might be
realistically designed and implemented.

Let us consider the initial pure atom-motion state
\begin{eqnarray}
| g \rangle \otimes \sum_{n} c_n | n \rangle
\end{eqnarray}
and tune our system to be selectively resonant within the subspace
$\{ | g, N_0 \rangle \leftrightarrow | e, N_0 + 1 \rangle \}$.
Then, we let it evolve for a time equivalent to a $\pi$-pulse,
$\Omega_{\rm eff} t \sqrt{N_0 + 1} = \pi$, in such a way that the
population of state $| g, N_0 \rangle$ is entirely transferred to
the state $| e, N_0 + 1 \rangle$
\begin{eqnarray}
| g \rangle \sum_{n} c_n | n \rangle \rightarrow | g \rangle
\sum_{n \neq N_0} c_n | n \rangle + c_{N_0} | e \rangle | N_0 + 1
\rangle .
\end{eqnarray}
Here, by measuring the population of the excited state $|e>$, we
project the motional state onto the Fock state $|N_0+1>$ with
probability $P_{e} = |c_{N_0}|^2$. In this way, we can generate
Fock state $| N_0 +1 \rangle$ out of any initial state containing
a finite population of the motional state $| N_0 \rangle$.

Similar results are obtained if the initial state is a thermal
state or any statistical mixture, as we will see. For example, let
us consider the ion initially in the ground state $| g \rangle$
and a precooled motional state, so as to have a finite
contribution of Fock state $| 1 \rangle$,
\begin{eqnarray}
| e \rangle \langle e | \otimes \sum_n p_{n} | n \rangle \langle n
| .
\end{eqnarray}
If we tune to resonance the subspace $\{ | g, 0 \rangle , | e, 1
\rangle \}$ the evolution corresponding to a $\pi$-pulse yields
\begin{eqnarray}
&& | g \rangle \langle g | \sum_{n} p_n | n \rangle \langle n |
\rightarrow  \nonumber  \\ [1.2ex]  | g \rangle \langle g | \!\!
\sum_{n \neq N_0} && \!\!\!\!\!\!\! p_n | n \rangle \langle n | +
p_{1} | e \rangle \langle e | | N_0 + 1 \rangle \langle N_0 + 1 |.
\end{eqnarray}
\begin{figure}[!t]
\includegraphics[width=60mm]{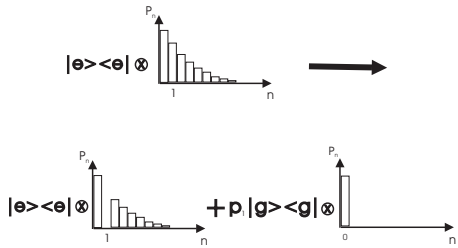}
\caption{\label{fig2} Sketch of a selective cooling mechanism.}
\end{figure}
Here, by measuring the internal state $|g>$, see Fig. 2, we
produce a "single-shot" cooling onto the motional ground state $|
0 \rangle$~\cite{Eschnercooling}. This unconventional cooling
mechanism requires a finite population of the Fock state $| 1
\rangle$ in the initial motional state in order to happen. From
that point of view, it could be used for resetting the ground
state in a previously cooled ion or ion chain.

Possible extensions of the notion of selectivity to other
subspaces, like $\{\left|g,N_0\right\rangle, \left| e,N_0 \pm k
\right\rangle\}$, or to subspaces involving many atoms, like
$\{\left|gg... , N_0 \right\rangle, \left| ee... ,N_0 \pm k
\right\rangle\}$,  and/or different collective vibrational modes,
are naturally expected~\cite{TwoIonSelectivity} but not developed
here. They would allow us to engineer arbitrary atom-motion
superposition states in a simplified
manner~\cite{ionsuperpositions}.

When we tuned to resonance the subspace $\{ | g, N_0 \rangle , |
e, N_0 + 1 \rangle \} $, under the selectivity condition ${\cal S}
\gg 1$, we showed that after a $\pi$-pulse
\begin{eqnarray}
P_{e} =  | c_{N_0} |^2 \equiv P_{N_0}  .
\end{eqnarray}
That implies that by measuring the population of the internal
excited state, $P_{e}$, we measure directly the population of a
preselected Fock state $| N_0 \rangle$ of an arbitrary and
initially unknown motional state. To our knowledge, it is the
first proposal for measuring directly a given motional population
without requiring the complete state reconstruction. If the
selectivity parameter ${\cal S}$ is not large enough, we could
repeat the selective procedure and set the experimental parameters
to put in resonance the subspace $\{ \left|g,N_0+1\right\rangle
\rightarrow \left| e,N_0+2\right\rangle \}$. The probability of
finding again the excited state $| e \rangle$ becomes closer to
$P_{N_0}$, yielding a continuously convergent measurement
technique. Combined with the possibility of displacing the
motional state~\cite{displacement}, this method permits a full
reconstruction of the associated Wigner function~\cite{Wigner}
\begin{eqnarray}
W(\alpha )= 2 \sum_{n}(-1)^{n}P_n(-\alpha),
\end{eqnarray}
where $P_n(\alpha)=\left\langle n\right| D(\alpha ) \rho
D^{-1}(\alpha )\left| n\right\rangle$ and $D(\alpha )$ is a
displacement operator. As known, the Wigner function contains the
same information as the density operator, and both are related
through a Fourier transform~\cite{Glauber}.

Selective addressing of subspaces and conditional dynamics are
interrelated concepts~\cite{Innsbruckselectivity}, so it should
not be a surprise that selectivity finds a natural environment in
tailored quantum state engineering, as well as in quantum-logic
devices. As an example, we propose an implementation of a
controlled-phase gate (CPG)~\cite{HarocheLogic} in the internal
states of two ions at arbitrary positions in a row of $N$.
Nevertheless, it is also possible to produce other gates, like
swap gates, with a reduced number of steps.

Let us consider the two internal levels of ion $j$, $\{ | g_j
\rangle , | e_j \rangle \}$, as the control qubit and the two
internal levels of ion $k$, $\{ | g_k \rangle , | e_k \rangle \}$,
as the target qubit. The protocol involves three steps where the
first and third consist in mapping forth and back, respectively,
the control qubit into the $\{ | 0 \rangle, | 1 \rangle \}$
motional states, and the second step consists in realizing a
selective CPG between the mapped motional state and the target
qubit. For the sake of simplicity, although its applicability is
general, we will illustrate the protocol assuming the initial pure
state
\begin{eqnarray}
( \alpha |g_j \rangle | g_k \rangle + \beta |g_j \rangle | e_k
\rangle + \gamma | e_j \rangle | g_k \rangle + \delta | e_j
\rangle | e_k \rangle ) | 0 \rangle
\end{eqnarray}
as a possible intermediate state in a certain computation.

The first step, that is the mapping of the target qubit $j$ onto
the motion, can be realized with a $\pi$-pulse of a selective
interaction in the subspace $\{ | g_j , 1 \rangle , | e_j , 0
\rangle \}$
\begin{eqnarray}
| g_j \rangle ( \alpha | 0 \rangle | g_k \rangle + \beta | 0
\rangle | e_k \rangle + \gamma | 1 \rangle | g_k \rangle + \delta
| 1 \rangle | e_k \rangle ) .
\end{eqnarray}

The second step, that is the selective CPG between the motional
state and ion $k$, can be realized by changing the sign of the
state $| g_k , 1 \rangle$ through a $2 \pi$-pulse in the
preselected subspace $\{ | e_k , 1 \rangle , | g_k , 2 \rangle \}$
\begin{eqnarray}
| g_j \rangle ( \alpha | 0 \rangle | g_k \rangle + \beta | 0
\rangle | e_k \rangle + \gamma | 1 \rangle | g_k \rangle - \delta
| 1 \rangle | e_k \rangle ) .
\end{eqnarray}

The final third step consists in mapping back the motional state
onto ion $j$ through a similar procedure
\begin{eqnarray}
( \alpha |g_j \rangle | g_k \rangle + \beta |g_j \rangle | e_k
\rangle + \gamma | e_j \rangle | g_k \rangle - \delta | e_j
\rangle | e_k \rangle ) | 0 \rangle .
\end{eqnarray}
This last equation reflects the implementation of a selective CPG
between the qubits in ions $j$ and $k$. It is noteworthy to
mention that this protocol is still valid if the initial motional
state is any superposition state $a | 0 \rangle + b | 1 \rangle$.
Unfortunately, we have not succeeded in finding a protocol robust
to any initial motional state.

We have considered the realistic implementation of selective
interactions in trapped ion systems. We have discussed a method
for generating Fock states, large or not, in the CM motion of a
single trapped ion or in a collective mode of an ion chain. We
showed that this scheme could be used as a {\it sui generis}
cooling device in precooled systems when the chosen Fock state is
the motional ground state. We demonstrated that selectivity offers
us the possibility of measuring distinctively the motional state
population and also, if required, its Wigner function in a
straightforward manner. We sketched that selective interactions,
when not enough accurate (${\cal S}$ not so large), can be applied
subsequently, yielding each time more accurate measurements. A
wide family of nonclassical states, linear superpositions,
entangled states, and quantum information devices could be
engineered by selectively tailoring the Hilbert space in this new
context. Possible implementations of quantum blockade and
turnstile devices in CQED and trapped ion systems are under
current research. We envisage also further investigation and
generalization of selective schemes in other physical systems like
optical lattices and atomic clouds.

This research was supported by the EU through the RESQ (Resources
for Quantum Information) project.


\begin{thebibliography}{}
\bibitem{comment} $\eta \sqrt{{\bar n}} \ll 1$, where ${\bar n}$ is
the average phonon number.
\end{thebibliography}
\end{document}